\documentclass[conference,10pt]{IEEEtran}

\usepackage{amsmath,amssymb,amsthm} 
\usepackage[dvips]{graphicx}
\usepackage{cite}
\usepackage{balance}

\newtheorem{proposition}{Proposition} 
\newtheorem{lemma}{Lemma} 
\newtheorem{theorem}{Theorem}

\begin{document}

%% Paper Title
%% You can use linebreaks \\ within to get better formatting as
%% desired. 
\title{A Generalization of Threshold Saturation: 
Application to Spatially Coupled BICM-ID} 
%%\title{ISIT 2013 Paper Template\\Please Capitalize Important Words in Title} 

%% Author names and affiliations:
%%
%% Avoiding spaces at the end of the author lines is not a problem with
%% conference papers because we don't use \thanks or \IEEEmembership.
%%
%% For several authors with only one affiliation:
%%
% \author{
%   \IEEEauthorblockN{Hui-Ting Chang and Stefan M.~Moser}
%   \IEEEauthorblockA{Department of Electrical and Computer Engineering\\
%     National Chiao Tung University (NCTU)\\
%     Hsinchu, Taiwan\\
%     Email: \{email-of-hui-ting,email-of-stefan\}@ieee.org} 
% }
%%
%% For up to three affiliations:
%%
\author{\IEEEauthorblockN{Keigo Takeuchi}
\IEEEauthorblockA{Dept.\ Commun.\ Eng.\ \& 
Informatics, University of Electro-Communications, 
Tokyo 182-8585, Japan}
\IEEEauthorblockA{Email: ktakeuchi@uec.ac.jp} 
}

\maketitle

%% Abstract: 
%% For the final version of the accepted paper, please make sure you
%% remove the comment "THIS PAPER IS ELIGIBLE FOR THE STUDENT PAPER
%% AWARD."
%%
\begin{abstract}
Spatial coupling was proved to improve the belief-propagation (BP) 
performance up to the maximum-a-posteriori (MAP) performance. 
This paper addresses an extended class of spatially coupled (SC) systems. 
A potential function is derived for characterizing a lower bound on the 
BP performance of the extended SC systems, and shown to be different from 
the potential for the conventional SC systems. This may imply that 
the BP performance for the extended SC systems does not coincide with the MAP 
performance for the corresponding uncoupled system. SC bit-interleaved coded 
modulation with iterative decoding (BICM-ID) is also investigated as 
an application of the extended SC systems. 
\end{abstract}

\section{Introduction}
Kudekar et al.~\cite{Kudekar11} proved that spatial coupling can improve 
the belief-propagation (BP) threshold up to the maximum-a-posteriori (MAP) 
threshold. Since the original proof of this {\em threshold saturation} was 
complicated, several simpler proofs have been developed in 
\cite{Yedla12,Donoho13,Takeuchi122,Kudekar12,Schlegel13}. 
In this paper we generalize the methodology in \cite{Takeuchi122} to 
characterize the BP performance for extended spatially-coupled (SC) systems. 

Consider the density-evolution (DE) equations of an extended SC system 
with the number of sections~$L$ and coupling width~$W$ 
\begin{equation} \label{DE_u_dis}
u_{l}(i+1) 
= \frac{1}{W^{d}}\sum_{\boldsymbol{w}_{d}\in\mathcal{W}^{d}}
\varphi(\boldsymbol{v}_{l+\boldsymbol{w}_{d}}(i)), 
\quad l\in\mathcal{L},  
\end{equation}
\begin{equation} \label{DE_v_dis} 
v_{l}(i) 
= \frac{1}{W^{\tilde{d}}}\sum_{\boldsymbol{w}_{\tilde{d}}\in\mathcal{W}^{\tilde{d}}}
\psi(\boldsymbol{u}_{l-\boldsymbol{w}_{\tilde{d}}}(i)), 
\quad l\in\{W-1,\ldots,L-1\}, 
\end{equation}
with $\mathcal{L}=\{0,\ldots,L-1\}$ and $\mathcal{W}=\{0,\ldots,W-1\}$. 
For notational convenience, 
we have used the notation $\boldsymbol{v}_{l+\boldsymbol{w}_{d}}(i)
=(v_{l+w_{1}}(i),\ldots,v_{l+w_{d}}(i))$ and 
$\boldsymbol{u}_{l-\boldsymbol{w}_{\tilde{d}}}(i)
=(u_{l-w_{1}}(i),\ldots,u_{l-w_{\tilde{d}}}(i))$, with  
$\boldsymbol{w}_{k}=(w_{1},\ldots,w_{k})$. 
The notation $l+\boldsymbol{w}$ should be interpreted as 
$(l,\ldots,l)+\boldsymbol{w}$. 
In (\ref{DE_u_dis}) and (\ref{DE_v_dis}), 
the state $(u_{l}(i),v_{l}(i))\in\mathcal{D}\times\tilde{\mathcal{D}}
\subset\mathbb{R}^{2}$ represents performance of a BP-based algorithm 
for section~$l\in\mathcal{L}$ in iteration~$i$. The two functions 
$\varphi:\tilde{\mathcal{D}}^{d}\rightarrow\mathcal{D}$ and 
$\psi:\mathcal{D}^{\tilde{d}}\rightarrow\tilde{\mathcal{D}}$ characterize the 
properties of the BP algorithm. The 
parameters $d+1$ and $\tilde{d}+1$ correspond to the degrees of check and 
variable nodes in low-density parity-check (LDPC) codes. In bit-interleaved 
coded modulation, $d+1$ is equal to the modulation rate, 
whereas $\tilde{d}=1$ is used. 

Without loss of generality, we postulate that larger variables 
$u_{l}(i)$ and $v_{l}(i)$ imply better performance. 
Let $(u_{\mathrm{opt}},v_{\mathrm{opt}})$ denote a fixed-point (FP) that has the 
largest $u$ among all FPs of the DE equations 
for the uncoupled case $W=1$. Thus, $(u_{\mathrm{opt}},v_{\mathrm{opt}})$ is a 
solution $(u,v)$ to the following FP equations:   
\begin{equation} \label{FP} 
u = \varphi_{0}(v), \quad v = \psi_{0}(u), 
\end{equation}
with $\varphi_{0}(v)=\varphi(v,\ldots,v)$ and $\psi_{0}(u)=\psi(u,\ldots,u)$. 
%We assume that the solution $(u_{\mathrm{opt}},v_{\mathrm{opt}})$ is asymptotically stable, so that $|\varphi_{0}'(v_{\mathrm{opt}})\psi_{0}'(u_{\mathrm{opt}})|<1$ must hold. 
The FP $(u_{\mathrm{opt}},v_{\mathrm{opt}})$ corresponds to the best possible 
performance achieved by the BP algorithm. 

We assume that $\varphi(v_{1},\ldots,v_{d})$ and $\psi(u_{1},\ldots,u_{\tilde{d}})$ 
are bounded, smooth\footnote{
In this paper, a function is said to be smooth if it is twice continuously 
differentiable. 
}, and strictly increasing in all arguments everywhere. 
The monotonicity implies that the performance of 
the BP algorithm improves monotonically. We impose the worst  
initial condition $u_{l}(0)=u_{\mathrm{min}}=\inf\mathcal{D}$ for all 
$l\in\mathcal{L}$ and the best boundary conditions $v_{l}(i)=v_{\mathrm{opt}}$ for 
any $l\notin\{W-1,\ldots,L-1\}$ and $i$. The aim of spatial coupling is to 
let the state $(u_{l}(i),v_{l}(i))$ converge toward 
$(u_{\mathrm{opt}},v_{\mathrm{opt}})$ for all sections $l\in\mathcal{L}$ after 
sufficiently many iterations via coupling. 

%We define a potential function that characterizes the BP threshold for the conventional SC system as 
%\begin{equation} \label{potential} 
%V(v) = -D(\psi_{0}^{-1}(v)\|v), 
%\end{equation}
%where the divergence $D(u\|v)$ in information geometry~\cite{Amari00} is given by 
%\begin{equation} \label{divergence} 
%D(u\|v) = \int_{u_{\mathrm{opt}}}^{u}\psi_{0}(\tilde{u})d\tilde{u} 
%+ \int_{v_{\mathrm{opt}}}^{v}\varphi_{0}(\tilde{v})d\tilde{v} 
%- uv + u_{\mathrm{opt}}v_{\mathrm{opt}}.  
%\end{equation}
%In (\ref{divergence}), we has selected an arbitrary constant of integration such that $V(v_{\mathrm{opt}})=0$. There is one-to-one correspondence between all stationary points of the potential~(\ref{potential}) and all solutions to the FP equations~(\ref{FP})~\cite{Yedla12}. More precisely, a FP $v$ is stable (respectively (resp.) unstable) if and only if $v$ is a stable (resp.\ unstable) solution of the potential~(\ref{potential}). 

We consider the continuum limit in which $L$ and $W$ tend to infinity while 
the ratio $\alpha=W/L$ is kept constant. The BP performance for 
the SC systems~(\ref{DE_u_dis}) and (\ref{DE_v_dis}) is characterized by 
a potential function for the uncoupled system 
\begin{IEEEeqnarray}{rl}  
V(u) =& \int\left\{
 u - \varphi_{0}(\psi_{0}(u))
\right\}\psi_{0}'(u) 
\nonumber \\ 
&\cdot\exp\left\{
 D(u;\psi) + D(\psi_{0}(u);\varphi) 
\right\}du, \label{true_potential}
\end{IEEEeqnarray}
with 
\begin{equation} \label{function_D} 
D(u;\psi)
= \int\frac{\bigtriangleup\psi(u)}{\psi'_{0}(u)}du 
- \ln\psi_{0}'(u), 
\end{equation} 
where the single-variate Laplacian $\bigtriangleup\psi(u)$ is given by 
$\bigtriangleup\psi(u)=\sum_{j}\partial^{2}\psi/\partial u_{j}^{2}
(u,\ldots,u)$. 
The goal of this paper is to prove the following statement: 
\begin{theorem} \label{theorem1}
Take the continuum limit $W=\alpha L\to\infty$, the infinite-iteration 
limit $i\to\infty$, and finally the limit $\alpha\to0$.  
If $u_{\mathrm{opt}}$ is the unique global stable solution (global minimizer) of 
the potential~(\ref{true_potential}), the state $(u_{l}(i),v_{l}(i))$ 
convergences to the target solution $(u_{\mathrm{opt}},v_{\mathrm{opt}})$ 
in the limits above.  
\end{theorem}
Theorem~\ref{theorem1} is a generalization of previous\footnote{
One should not regard that only degree~$d+1=2$ was considered in 
\cite{Yedla12,Donoho13,Schlegel13,Takeuchi122,Kudekar12}, 
although $d+1$ and $\tilde{d}+1$ are interpreted as the degrees of nodes for 
factor graphs in this paper. Since $\varphi$ and $\psi$ have a special 
structure for SC LDPC codes, 
the DE equations~(\ref{DE_u_dis}) and (\ref{DE_v_dis}) reduce to those 
with $d=\tilde{d}=1$.      
} works~\cite{Yedla12,Donoho13,Schlegel13,Takeuchi122,Kudekar12} for 
$d=\tilde{d}=1$, and implies that the qualitative {\em shape} of the 
potential~(\ref{true_potential}) determines whether the BP algorithm can 
achieve the best possible performance point $(u_{\mathrm{opt}},v_{\mathrm{opt}})$, 
whereas the uniqueness of solutions to the potential~(\ref{true_potential}) 
does for the uncoupled case $W=1$. 
The potential~(\ref{true_potential}) reduces to the 
conventional one defined in \cite{Yedla12} for $d=\tilde{d}=1$,  
and coincides with the conventional one for $d,\tilde{d}>1$ if  
$D(u;\psi)+D(\psi_{0}(u);\varphi)$ is independent of $u$. 
The latter observation may imply that for $d,\tilde{d}>1$ the BP threshold 
does not coincide with the MAP threshold for the corresponding uncoupled 
system, since the potential for $d=\tilde{d}=1$ is used to characterize 
the MAP threshold. 

Theorem~\ref{theorem1} is useful when no analytical formulas of the 
multi-variate functions~$\varphi$ and $\psi$ 
in (\ref{DE_u_dis}) and (\ref{DE_v_dis}) are available. In this case, 
the cost for calculating the two functions numerically increases 
exponentially as $d$ and $\tilde{d}$ grow. Theorem~\ref{theorem1} implies that 
we can know the BP performance just by estimating six single-variate 
functions $\varphi_{0}$, $\varphi_{0}'$, $\bigtriangleup \varphi$, $\psi_{0}$, 
$\psi_{0}'$, and $\bigtriangleup\psi$ via numerical integration or sampling. 

This paper is organized as follows: In Section~\ref{sec2} we shall present 
an application of Theorem~\ref{theorem1} to SC bit-interleaved coded 
modulation with iterative decoding (BICM-ID). 
Theorem~\ref{theorem1} is proved in Section~\ref{sec3}.  

\section{Application} \label{sec2} 
\subsection{Spatially Coupled BICM-ID}
We consider a BICM-ID scheme with SC interleaving~\cite{Takeuchi13}. 
One transmission consists of $W-1$ binary training sequences of length~$M$ and 
of $L-(W-1)$ binary codewords of length $M$. The training sequences are 
utilized to anchor the performance of the system at the boundaries to the best 
performance $v_{\mathrm{opt}}$. After SC interleaving of length~$LM$, 
the obtained binary sequences are mapped to $LM/Q$ data symbols, with $Q$ 
denoting the modulation rate. The data symbols are transmitted through a 
memoryless time-invariant communication channel, and detected with iterative 
decoding at the receiver side~\cite{Li02}.  

We shall review a construction of SC interleaving~\cite{Takeuchi13}. 
Let $\{\pi_{l}^{\mathrm{in}}:l\in\mathcal{L}\}$ and 
$\{\pi_{l}^{\mathrm{out}}:l\in\mathcal{L}\}$ denote $2L$ 
independent random interleavers of length~$M$ that are bijections from 
$\mathcal{M}=\{0,\ldots,M-1\}$ onto $\mathcal{M}$.  
An SC interleaver $\pi(m,l)$ is a bijection from $\mathcal{M}\times\mathcal{L}$ 
onto $\mathcal{M}\times\mathcal{L}$ that maps $m$th bit in section~$l$ to 
the pair $\pi(m,l)$ of bit and section indices, 
\begin{equation}
\pi(m,l) = (\pi_{l'}^{\mathrm{out}}(\pi_{l}^{\mathrm{in}}(m)),l'), 
\quad l'=(l-(\pi_{l}^{\mathrm{in}}(m))_{W})_{L}, 
\end{equation}
where $(i)_{n}=i\mod n\in\{0,\ldots,n-1\}$ denotes the remainder for the 
division of $i\in\mathbb{Z}$ by $n\in\mathbb{N}$. From the construction, 
$M$ bits in section~$l$ are sent to sections $\{l,\ldots,(l-(W-1))_{L}\}$ with 
uniform frequency when $M$ is a multiple of $W$. As a result,  
each bit in section~$l'$ at the output side originates from a bit in 
sections $\{l',\ldots,(l'+W-1)_{L}\}$ with equal probability.  
These properties result in the DE equations~(\ref{DE_u_dis}) and 
(\ref{DE_v_dis}) with $d=Q-1$ and $\tilde{d}=1$ when $M$ tends to infinity. 
Minus one is because iterative decoding is based on 
extrinsic feedback information. 

\subsection{EXIT Chart Analysis} 
Let us consider a mathematical model based on erasure extrinsic 
channels~\cite{Ashikhmin04} that approximates the dynamical properties of 
the SC BICM-ID scheme in the limit $M\to\infty$. 
The distributions of messages passed between the demodulator and the decoder 
are very complicated in general. This may be regarded as if bits were sent 
through {\em extrinsic} channels subject to very complicated noise. 
In order to approximate the message distributions with tractable one-parameter 
distributions, the extrinsic channels are replaced by binary erasure channels 
(BECs) with the same input-output mutual information as the original one. 
Consequently, it is sufficient to evaluate the dynamics of the mutual 
information, instead of that of complicated distributions.  

\begin{figure}[t]
\begin{center}
\includegraphics[width=\hsize]{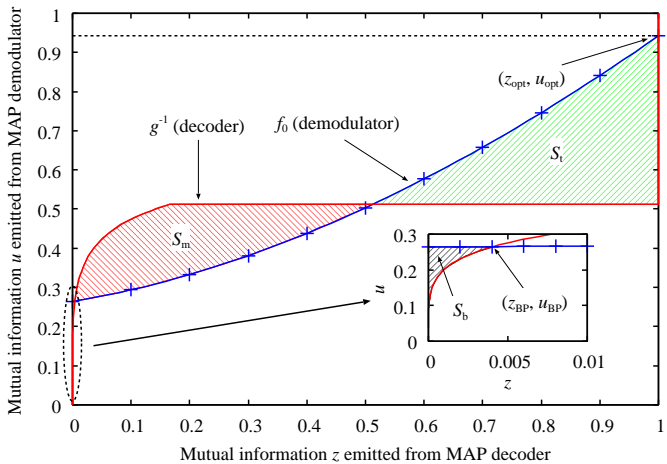}
\end{center}
\caption{
EXIT chart of the conventional BICM-ID with $16$-QAM 
for erasure extrinsic and AWGN communication channels with an SNR of $5.76$~dB. 
}
\label{fig1}
\end{figure}

Under these assumptions, the DE equations for the SC BICM-ID scheme are given 
by (\ref{DE_u_dis}) and (\ref{DE_v_dis}) with $d=Q-1$ and $\tilde{d}=1$. 
The variable $u_{l}(i)\in[0,1]$ corresponds to the mutual 
information emitted from the demodulator for section~$l$ in iteration~$i$, 
whereas $v_{l}(i)\in[0,1]$ is the average mutual 
information that is the input to the decoder in section~$l$. 
The identity function $\psi(u)=u$ with $\tilde{d}=1$ is used. 
Let $f(I_{1},\ldots,I_{Q-1})$ and $g(I)$ denote the extrinsic information 
transfer (EXIT) functions for the MAP demodulator and the MAP decoder, 
respectively. The function $\varphi$\footnote{
Although $g(x)$ is a discontinuous non-decreasing function as shown in 
Fig.~\ref{fig1}, we can use Theorem~\ref{theorem1} by considering a sequence 
$\{g_{n}(x)\}_{n=1}^{\infty}$ of smooth increasing functions that converges 
toward $g(x)$ as $n\to\infty$.  
} is given by 
$\varphi(v_{1},\ldots,v_{Q-1})=f(g(v_{1}),\ldots,g(v_{Q-1}))$.  
Introducing a variable $z_{l}(i)$ that represents the mutual 
information emitted from the decoder for section~$l$ in iteration~$i$, 
we find that the DE equations~(\ref{DE_u_dis}) and (\ref{DE_v_dis}) are 
represented as 
\begin{equation}
u_{l}(i+1) = \frac{1}{W^{Q-1}}\sum_{\boldsymbol{w}_{Q-1}\in\mathcal{W}^{Q-1}}
f(\boldsymbol{z}_{l+\boldsymbol{w}_{Q-1}}(i)), 
\quad l\in\mathcal{L},  
\end{equation}
\begin{equation}
z_{l}(i) = g\left(
 \frac{1}{W}\sum_{w=0}^{W-1}u_{l-w}(i)
\right), 
\quad l\in\{W-1,\ldots,L-1\}, 
\end{equation}
where $\boldsymbol{z}_{l+\boldsymbol{w}_{Q-1}}(i)$ is defined in the same manner 
as for $\boldsymbol{v}_{l+\boldsymbol{w}_{d}}(i)$. 

Figure~\ref{fig1} shows the EXIT chart for the uncoupled case $W=1$. 
We used $(3,6)$-LDPC codes, quadrature amplitude modulation (QAM) with $Q=4$ 
and a symbol mapping proposed in \cite{Tan05}, and the additive white 
Gaussian noise (AWGN) communication channel. We utilized an analytical 
expression of $g$ for the MAP decoder~\cite{Measson08}. 
In practice, one may use the corresponding SC LDPC code based on BP decoding. 
We selected a signal-to-noise ratio (SNR) of $5.76$~dB that is slightly larger 
than the minimum of SNRs such that $u_{\mathrm{opt}}$ is the unique global 
stable solution of the potential~(\ref{true_potential}) in 
Theorem~\ref{theorem1}, so that the SC BICM-ID scheme can approach the 
target point $(z_{\mathrm{opt}},u_{\mathrm{opt}})=(1,f_{0}(1))$, with 
$f_{0}(I)=f(I,\ldots,I)$.   
We find that the FP equations~(\ref{FP}) have two stable solutions. 
One stable solution is the target solution 
$(z_{\mathrm{opt}},u_{\mathrm{opt}})=(1,f_{0}(1))$,  
and the other stable solution $(z_{\mathrm{BP}},u_{\mathrm{BP}})$ is a FP to which 
the BP algorithm converges for the uncoupled case $W=1$. These observations 
imply that the conventional BICM-ID scheme cannot approach the target point 
$(1,f_{0}(1))$ in this case, whereas the SC BICM-ID scheme can. 

It is possible to understand the performance of the SC BICM-ID 
scheme from the EXIT chart for the {\em un}coupled case. 
Let $S_{\mathrm{t}}$, $S_{\mathrm{m}}$, and $S_{\mathrm{b}}$ denote the three areas 
enclosed by the two curves in Fig.~\ref{fig1} from top to bottom. 
From the area theorems for the MAP decoder~\cite{Ashikhmin04}, 
\cite[Corollary 5.1]{Fabregas08} and the MAP 
demodulator~\cite[Corollary 5.2]{Fabregas08}, it is straightforward to 
find the relationship between the areas and the rate loss from 
the coded modulation (CM) capacity $C_{\mathrm{CM}}$~\cite{Fabregas08}
\begin{equation} \label{rate_loss} 
C_{\mathrm{CM}} - Qr = QS_{\mathrm{b}} + Q(S_{\mathrm{t}}-S_{\mathrm{m}}),  
\end{equation}    
with $r$ denoting the code rate. 

Expression~(\ref{rate_loss}) implies that 
the rate loss from the CM capacity is characterized by $S_{\mathrm{b}}$ 
and $S_{\mathrm{t}}-S_{\mathrm{m}}$. We note that $S_{\mathrm{t}}=S_{\mathrm{m}}$ holds 
at the BP threshold for conventional SC systems~\cite{Kudekar12}. 
Since $S_{\mathrm{b}}$ is much smaller 
than $S_{\mathrm{t}}-S_{\mathrm{m}}$ in Fig.~\ref{fig1}, the gap between 
$S_{\mathrm{t}}$ and $S_{\mathrm{m}}$ is a dominant factor for the rate loss. 
In fact, the SNR that $S_{\mathrm{t}}=S_{\mathrm{m}}$ holds is approximately 
$5.29$~dB. Furthermore, the SNR corresponding to the CM capacity is 
approximately given by $5.12$~dB. Since an SNR of $5.76$~dB was considered 
in Fig.~\ref{fig1}, the losses due to $S_{\mathrm{b}}$ and 
$S_{\mathrm{t}}-S_{\mathrm{m}}$ are given by $5.29-5.12=0.17$~dB and 
$5.76-5.29=0.47$~dB, respectively, if $S_{\mathrm{b}}$ is assumed to be 
identical for the two SNRs $5.29$~dB and $5.76$~dB.

%If $\varphi$ for the SC BICM-ID scheme were Laplacian-free, $S_{\mathrm{t}}=S_{\mathrm{m}}$ would hold~\cite{Kudekar12} when the SNR is equal to the improved BP threshold. Thus, expression~(\ref{rate_loss}) would imply that $QS_{\mathrm{b}}$ corresponds to the rate loss from the CM capacity. The author is attempting to estimate the Laplacian of the EXIT function $f$ for the MAP demodulator numerically, and to use Theorem~\ref{theorem1} to specify the relationship between $\mathcal{S}_{\mathrm{m}}$ and $\mathcal{S}_{\mathrm{t}}$ for the case where the SNR is equal to a (possibly tight) lower bound on the improved BP threshold. 

\section{Proof of Theorem~\ref{theorem1}} \label{sec3} 
\subsection{Sketch}
The proof of Theorem~\ref{theorem1} is a generalization of that in 
\cite{Takeuchi122}. 
We shall define a partial differential equation that characterizes the FPs to 
the DE equations~(\ref{DE_u_dis}) and (\ref{DE_v_dis}).    
%\begin{equation} \label{energy_functional} 
%F_{2}(v;\alpha) = \int_{-1}^{1}\left[
% V(v(x)) + \frac{\alpha^{2}}{2}A(v(x))\left(
%  \frac{dv}{dx}
% \right)^{2}
%\right]dx, 
%\end{equation}
%\begin{equation}
%A(v) = \frac{1}{3}\varphi_{0}'(v). 
%\end{equation}
\begin{equation}
\frac{\partial\tilde{u}}{\partial t}(x,t) 
= - \tilde{u}(x,t) + \tilde{\mathfrak{L}}[\tilde{u}(\cdot,t)](x), 
\label{differential_system_u}
\end{equation}
where the differential operator $\tilde{\mathfrak{L}}[\tilde{u}](x)$ 
for any smooth function $\tilde{u}(x)$ on $[-1,1]$ is defined as 
\begin{equation} \label{differential_operator} 
\tilde{\mathfrak{L}}[\tilde{u}](x) 
= \varphi_{0}(\psi_{0}(\tilde{u}))  
+ \alpha^{2}\left[
 A(\tilde{u})\left(
  \frac{d\tilde{u}}{dx} 
 \right)^{2}
 + B(\tilde{u})\frac{d^{2}\tilde{u}}{dx^{2}} 
\right], 
\end{equation}
with $\tilde{u}$ denoting an abbreviation of $\tilde{u}(x)$. 
In (\ref{differential_operator}), 
the two functions $A(\tilde{u})$ and $B(\tilde{u})$ are given by 
\begin{equation} \label{function_A}
A(\tilde{u}) = \frac{\varphi_{0}'(\psi_{0}(\tilde{u}))}{6}\left(
 \bigtriangleup\psi(\tilde{u})
 + \frac{\bigtriangleup\varphi(\psi_{0}(\tilde{u}))\psi_{0}'(\tilde{u})^{2}}
 {\varphi_{0}'(\psi_{0}(\tilde{u}))}
 + \psi_{0}''(\tilde{u})
\right), 
\end{equation}
\begin{equation} \label{function_B} 
B(\tilde{u}) 
= \frac{1}{3}\varphi_{0}'(\psi_{0}(\tilde{u}))\psi_{0}'(\tilde{u}) > 0,
\end{equation}
respectively. 
We impose the boundary condition $\tilde{u}(\pm 1,t)=u_{\mathrm{opt}}$. 
Furthermore, an initial condition 
$\tilde{u}(x,0)=\tilde{u}_{\mathrm{init}}(x)$ is imposed with a smooth 
function $\tilde{u}_{\mathrm{init}}(x)$. 

\begin{theorem} \label{theorem2}
There is some initial function $\tilde{u}_{\mathrm{init}}(x)$ such that 
\begin{equation} \label{difference} 
\lim_{\alpha\to0}\lim_{i\to\infty}\lim_{W=\alpha L\to\infty}\frac{1}{L}
\sum_{l\in\mathcal{L}}\left|
 u_{l}(i) - \tilde{u}\left(
  \frac{2l}{L}-1
 \right)
\right|=0, 
\end{equation}
with $\tilde{u}(x)=\lim_{t\to\infty}\tilde{u}(x,t)$. 
\end{theorem}
\begin{IEEEproof} 
See Section~\ref{sec3B}. 
\end{IEEEproof}

From Theorem~\ref{theorem2}, it is sufficient to analyze the stationary 
solution $\tilde{u}(x)$ to the partial differential 
equation~(\ref{differential_system_u}), which satisfies 
\begin{equation} \label{stationary_system} 
\tilde{u} - \varphi_{0}(\psi_{0}(\tilde{u})) 
= \alpha^{2}\left[
 A(\tilde{u})\left(
  \frac{d\tilde{u}}{dx} 
 \right)^{2}
 + B(\tilde{u})\frac{d^{2}\tilde{u}}{dx^{2}} 
\right], 
\end{equation}
with the boundary condition $\tilde{u}(\pm1)=u_{\mathrm{opt}}$.  
Theorem~\ref{theorem1} follows immediately from the following theorem. 

\begin{theorem} \label{theorem3} 
If and only if $u_{\mathrm{opt}}$ is the unique global stable solution of 
the potential~(\ref{true_potential}), the uniform solution 
$\tilde{u}(x)=u_{\mathrm{opt}}$ is the unique solution to the boundary-value 
problem~(\ref{stationary_system}) in the limit~$\alpha\to0$. 
\end{theorem}
\begin{IEEEproof}
We first present a coordinate system that simplifies the representation of 
the differential system~(\ref{stationary_system}). 
Let us define the change of variables $y=f(\tilde{u})$ by 
\begin{equation} \label{coordinate_system}
f(\tilde{u}) = \int e^{C(\tilde{u})}d\tilde{u}, 
\end{equation}
with 
\begin{equation} \label{function_C} 
C(\tilde{u}) = \int\frac{A(\tilde{u})}{B(\tilde{u})}d\tilde{u}. 
\end{equation}
Calculating $d^{2}y/dx^{2}$ with the chain rule for partial derivative yields 
\begin{equation}
\frac{d^{2}y}{dx^{2}} 
= \frac{e^{C(\tilde{u})}}{B(\tilde{u})}\left[
 A(\tilde{u})\left(
  \frac{d\tilde{u}}{dx}
 \right)^{2}
 + B(\tilde{u})\frac{d^{2}\tilde{u}}{dx^{2}} 
\right]. 
\end{equation}
Thus, the differential equation~(\ref{stationary_system}) 
for stationary solutions reduces to 
\begin{equation} \label{Newton_equation} 
\alpha^{2}\frac{d^{2}y}{dx^{2}}(x)= \tilde{V}'(y(x)), 
\end{equation}
where the derivative of a potential $\tilde{V}(y)$ is given by 
\begin{equation} \label{potential_deriv} 
\tilde{V}'(y) = \left\{
 \tilde{u} - \varphi_{0}(\psi_{0}(\tilde{u})) 
\right\}\frac{e^{C(\tilde{u})}}{B(\tilde{u})}, 
\end{equation}
with $\tilde{u}=f^{-1}(y)$. It is straightforward to confirm that 
$\tilde{V}'(y)=0$ if and only if $\tilde{u}=f^{-1}(y)$ is a 
solution to the FP equation $\tilde{u}=\varphi_{0}(\psi_{0}(\tilde{u}))$ 
obtained from (\ref{FP}) for the uncoupled system. In particular, 
any stable solution to the potential $\tilde{V}$ corresponds to a stable 
FP to (\ref{FP}). Thus, $y_{\mathrm{opt}}=f(u_{\mathrm{opt}})$ is a stable solution 
to $\tilde{V}(y)$. 

It can be proved that the uniform solution $y(x)=y_{\mathrm{opt}}$ is the unique 
solution to the boundary-value problem~(\ref{Newton_equation}) with 
$y(\pm1)=y_{\mathrm{opt}}$ if and only if $y_{\mathrm{opt}}$ is the 
unique global stable solution of $\tilde{V}(y)$.  
Hassani et al.~\cite{Hassani12} presented an intuitive explanation of this  
statement based on classical mechanics. See \cite{Takeuchi121,Takeuchi122} 
for a rigorous proof based on the intuition. 

Let us prove that the potential $\tilde{V}(y)$ 
defined via (\ref{potential_deriv}) is equivalent to (\ref{true_potential}). 
By definition, we use (\ref{coordinate_system}) and (\ref{potential_deriv}) 
to obtain 
\begin{equation} \label{potential} 
\tilde{V}(y) = \int \tilde{V}'(y)dy 
= \int\left\{
 \tilde{u} - \varphi_{0}(\psi_{0}(\tilde{u})) 
\right\}\frac{e^{2C(\tilde{u})}}{B(\tilde{u})}d\tilde{u}. 
\end{equation}
We calculate (\ref{function_C}) with (\ref{function_A}) and 
(\ref{function_B}) to obtain 
\begin{IEEEeqnarray}{rl}
2C(\tilde{u})
=& \int\left\{
 \frac{\bigtriangleup\psi(\tilde{u})}{\psi_{0}'(\tilde{u})}
 + \frac{\bigtriangleup\varphi(\psi_{0}(\tilde{u}))\psi_{0}'(\tilde{u})}
 {\varphi_{0}'(\psi_{0}(\tilde{u}))}
 + \frac{\psi_{0}''(\tilde{u})}{\psi_{0}'(\tilde{u})} 
\right\}d\tilde{u} \nonumber \\ 
=& D(\tilde{u};\psi) + D(\psi_{0}(\tilde{u});\varphi) 
+ \ln B(\tilde{u}) + \ln\psi_{0}'(\tilde{u}),  
\end{IEEEeqnarray}
with (\ref{function_D}). 
Substituting this expression into (\ref{potential}), we arrive at 
$\tilde{V}(y)=V(f^{-1}(y))$ given by (\ref{true_potential}). 
This implies that Theorem~\ref{theorem3} holds.  
\end{IEEEproof}

\subsection{Proof of Theorem~\ref{theorem2}} \label{sec3B}
We first confirm that the DE equations~(\ref{DE_u_dis}) 
and (\ref{DE_v_dis}) are convergent as $i\to\infty$. 
\begin{lemma} \label{lemma1}
For any $i$ and $l$, $u_{l}(i)\leq u_{l}(i+1)$ holds. 
\end{lemma}
\begin{IEEEproof}
The proof is by induction. The initial condition $u_{l}(0)=u_{\mathrm{min}}$ 
implies $u_{l}(i)\leq u_{l}(i+1)$ for $i=0$. Suppose that 
$u_{l}(i)\leq u_{l}(i+1)$ holds for some $i$. Since 
$\psi$ is increasing in all arguments, from (\ref{DE_v_dis}) we obtain 
\begin{IEEEeqnarray}{rl}
&v_{l}(i+1) - v_{l}(i) \nonumber \\ 
=& \frac{1}{W^{\tilde{d}}}\sum_{\boldsymbol{w}_{\tilde{d}}\in\mathcal{W}^{\tilde{d}}}
[\psi(\boldsymbol{u}_{l-\boldsymbol{w}_{\tilde{d}}}(i+1))
- \psi(\boldsymbol{u}_{l-\boldsymbol{w}_{\tilde{d}}}(i))] 
\nonumber \\ 
\geq&0, 
\label{monotonicity} 
\end{IEEEeqnarray}
for $l\in\{W-1,\ldots,L-1\}$. Combining (\ref{monotonicity}) and 
the boundary condition $v_{l}(i)=v_{l}(i+1)=v_{\mathrm{opt}}$ 
for $l\notin\{W-1,\ldots,L-1\}$, we obtain $v_{l}(i)\leq v_{l}(i+1)$ for 
all $l$. Repeating the same argument for (\ref{DE_u_dis}), 
we arrive at $u_{l}(i+1)\leq u_{l}(i+2)$ for all $l$. 
By induction, Lemma~\ref{lemma1} holds. 
\end{IEEEproof}

Theorem~\ref{theorem2} is proved as follows: We first take the continuum 
limit to reduce the DE equations~(\ref{DE_u_dis}) and (\ref{DE_v_dis}) 
to integral systems. Then, we shall derive the differential 
system~(\ref{differential_system_u}) by expanding the integral systems 
with respect to $\alpha$. Finally, we investigate the relationship between 
stationary solutions for the integral and differential systems 
as $\alpha\to0$.  

Let us define the integral systems as 
\begin{equation} \label{DE_u_cont}
u(x,i+1) = 
\frac{1}{(2\alpha)^{d}}\int_{[-\alpha,\alpha]^{d}}
\varphi(\boldsymbol{v}(x+\boldsymbol{\omega}_{d},i))d\boldsymbol{\omega}_{d},  
\;\; |x|\leq1,
\end{equation}
\begin{equation} \label{DE_v_cont} 
v(x,i) = 
\frac{1}{(2\alpha)^{\tilde{d}}}
\int_{[-\alpha,\alpha]^{\tilde{d}}}
\psi(\boldsymbol{u}(x-\boldsymbol{\omega}_{\tilde{d}},i))
d\boldsymbol{\omega}_{\tilde{d}},  
\;\; |x|\leq1-\alpha,
\end{equation}
where we have introduced the notation 
$\boldsymbol{v}(x+\boldsymbol{\omega}_{d},i)=(v(x+\omega_{1},i),\ldots,
v(x+\omega_{d},i))$ and $\boldsymbol{u}(x-\boldsymbol{\omega}_{\tilde{d}},i)
=(u(x-\omega_{1},i),\ldots,u(x-\omega_{\tilde{d}},i))$, with 
$\boldsymbol{\omega}_{k}=(\omega_{1},\ldots,\omega_{k})$. 
We impose the initial condition $u(x,0)=u_{\mathrm{min}}$ for all $|x|\leq1$. 
Furthermore, the boundary condition $v(x,i)=v_{\mathrm{opt}}$ is imposed 
for all $|x|>1-\alpha$ and all $i$. 
Since the two functions $\varphi$ and $\psi$ have been assumed to be bounded 
and continuous, the integral systems~(\ref{DE_u_cont}) and (\ref{DE_v_cont}) 
are well defined for any $i$. 

The integral systems~(\ref{DE_u_cont}) and (\ref{DE_v_cont}) define   
a recursive formula $u(x,i+1)=\mathfrak{L}[u(\cdot,i)](x)$ with respect to 
$u(x,i)$, in which the operator $\mathfrak{L}$ is a counterpart 
of the differential operator $\tilde{\mathfrak{L}}$ given by 
(\ref{differential_operator}). 

\begin{lemma} \label{lemma3}
\begin{enumerate}
\item For any $x$, $i$, and any $\alpha>0$, 
\begin{equation}
u(x,i) \leq u(x,i+1), 
\quad v(x,i) \leq v(x,i+1). 
\end{equation}
\item For any $i$ and $l\in\mathcal{L}$, 
\begin{equation}
\lim_{W=\alpha L\to\infty}\frac{1}{L}\sum_{l\in\mathcal{L}}\left|
 u_{l}(i) - u\left(
  \frac{2l}{L} - 1,i
 \right)
\right|=0, 
\end{equation}
\begin{equation}
\lim_{W=\alpha L\to\infty}\frac{1}{L}\sum_{l\in\mathcal{L}}\left|
 v_{l}(i) - v\left(
  \frac{2l}{L} - 1-\alpha,i
 \right)
\right|=0. 
\end{equation}
\item For any $i$, $u(x,i)$ is even, continuous on $[-1,1]$, and 
smooth on $(-1,1)-\{\pm(1-2\alpha)\}$. 
The stationary solution $u(x)=\lim_{i\to\infty}u(x,i)$ also has the same 
properties as $u(x,i)$. 
\end{enumerate}
\end{lemma}
\begin{IEEEproof}
See \cite{Takeuchi122}. 
\end{IEEEproof}
From the last property of Lemma~\ref{lemma3}, there exists some smooth 
initial function $\tilde{u}_{\mathrm{init}}(x)$ that is sufficiently close to 
the FP $u(x)$ of the integral systems. We impose the initial condition 
$\tilde{u}(x,0)=\tilde{u}_{\mathrm{init}}(x)$ for the differential 
system~(\ref{differential_system_u}) with such an initial function.

We next summarize several properties of the partial differential 
equation~(\ref{differential_system_u}). 

\begin{lemma} \label{stability} 
For any $\epsilon>0$ and $x\in[-1,1]$, there exist some $t_{0}>0$ and 
stationary solution $\tilde{u}(x)$ such that 
\begin{equation}
|\tilde{u}(x,t) - \tilde{u}(x)|<\epsilon, 
\end{equation}
for all $t\geq t_{0}$ and $\alpha>0$. 
\end{lemma}
\begin{IEEEproof}
See \cite{Takeuchi122}. 
\end{IEEEproof}

\begin{proposition} \label{proposition1}
Suppose that $u(x)$ is any smooth function on $[-1,1]$. 
For any $\epsilon>0$, there exists some $\alpha_{0}>0$ such that 
\begin{equation} \label{operator_difference} 
\int_{-1}^{1}\left|
 \mathfrak{L}[u](x) - \tilde{\mathfrak{L}}[u](x)  
\right|dx < \epsilon, 
\end{equation} 
for all $\alpha\in(0,\alpha_{0})$. 
\end{proposition}
\begin{IEEEproof}
Let $\mathcal{X}=(-(1-2\alpha),1-2\alpha)$ and 
$\bar{\mathcal{X}}=[-1,1]\backslash\mathcal{X}$ denote the bulk and 
boundary regions, respectively. We decompose the 
integral~(\ref{operator_difference}) into 
two parts. 
\begin{IEEEeqnarray}{rl} 
&\int_{-1}^{1}\left|
 \mathfrak{L}[u](x) - \tilde{\mathfrak{L}}[u](x)  
\right|dx 
\nonumber \\
=& \int_{\mathcal{X}}\left|
 \mathfrak{L}[u](x) - \tilde{\mathfrak{L}}[u](x)  
\right|dx
+ \int_{\bar{\mathcal{X}}}\left|
 \mathfrak{L}[u](x) - \tilde{\mathfrak{L}}[u](x)  
\right|dx. 
\nonumber \\ 
\end{IEEEeqnarray}
It is straightforward to show that the second term tends to zero as 
$\alpha\to0$. Thus, we focus on the first term. 

To complete the proof of Proposition~\ref{proposition1}, 
it is sufficient to prove that the integrand 
$|\mathfrak{L}[u](x) - \tilde{\mathfrak{L}}[u](x)|$ 
converges to zero as $\alpha\to0$ for all $x\in\mathcal{X}$ in 
the bulk region~\cite{Takeuchi122}. 
Since $u(x)$ is smooth, we can expand $\mathfrak{L}[u](x)$ 
with respect to $\alpha$ up to the second order. Expanding the integrand 
in (\ref{DE_u_cont}) with respect to $\boldsymbol{\omega}_{d}$ yields  
\begin{equation} \label{DE_u_approx} 
\mathfrak{L}[u](x) 
= \varphi_{0}(v) 
+ \frac{\alpha^{2}}{6}\left\{
 \bigtriangleup\varphi(v)\left(
  \frac{dv}{dx} 
 \right)^{2} + \varphi_{0}'(v)\frac{d^{2}v}{dx^{2}} 
\right\} 
+ o(\alpha^{2}), 
\end{equation} 
where $v$ is given by the right-hand side (RHS) of (\ref{DE_v_cont}) with 
$u(x,i)=u(x)$. Similarly, expanding $v$ with respect to 
$\alpha$ gives 
\begin{equation} \label{DE_v_approx} 
v = \psi_{0}(u) + \frac{\alpha^{2}}{6}\left[
 \bigtriangleup\psi(u)\left(
  \frac{du}{dx} 
 \right)^{2} + \psi_{0}'(u)\frac{d^{2}u}{dx^{2}}  
\right] + o(\alpha^{2}), 
\end{equation}
where $u$ is an abbreviation of $u(x)$.  
Substituting (\ref{DE_v_approx}) into (\ref{DE_u_approx}) and expanding 
the obtained formula with respect to $\alpha$, 
we obtain $\mathfrak{L}[u](x)=\tilde{\mathfrak{L}}[u](x)+o(\alpha^{2})$, 
given by (\ref{differential_operator}). 
\end{IEEEproof}

\begin{lemma} \label{lemma4} 
For any $t_{0}>0$ and $\epsilon>0$, there exists some $\alpha_{0}>0$ such that 
\begin{equation}
\int_{-1}^{1}|\tilde{u}(x,t_{0}) - \tilde{u}(x,0)|dx<\epsilon, 
\end{equation}
for all $\alpha\in(0,\alpha_{0})$. 
\end{lemma}
\begin{IEEEproof}
The proof is based on Proposition~\ref{proposition1}. See \cite{Takeuchi122} 
for the details. 
\end{IEEEproof}

\balance 

We are ready to prove Theorem~\ref{theorem2}.
\begin{IEEEproof}[Proof of Theorem~\ref{theorem2}]
Let $x_{l}=(2l/L)-1$. 
Lemma~\ref{lemma1} and the first property of Lemma~\ref{lemma3} imply that, 
for any $l\in\mathcal{L}$, $\alpha>0$, and any $\epsilon>0$, 
there exists some $I\in\mathbb{N}$ such that 
\begin{equation}
|u_{l}(i) - u_{l}(I)|<\epsilon, 
\quad |u(x_{l},i) - u(x_{l})|<\epsilon, 
\end{equation} 
for all $i\geq I$, 
with $u(x)=\lim_{i\to\infty}u(x,i)$ denoting the stationary solution to 
the integral systems~(\ref{DE_u_cont}) and (\ref{DE_v_cont}). 
With this number $I$ of iterations 
we use the triangle inequality for the left-hand side (LHS) 
of (\ref{difference}) to obtain 
\begin{IEEEeqnarray}{rl}
\frac{1}{L}\sum_{l\in\mathcal{L}}|u_{l}(i) - \tilde{u}(x_{l})| 
<& \frac{1}{L}\sum_{l\in\mathcal{L}}|u_{l}(I) - u(x_{l},I)|  
\nonumber \\ 
+& \frac{1}{L}\sum_{l\in\mathcal{L}}|u(x_{l}) - \tilde{u}(x_{l})|  
+ 2\epsilon,   \label{upper_bound}
\end{IEEEeqnarray}
for all $i\geq I$.  

From the second property of Lemma~\ref{lemma3}, we find that the first term 
on the upper bound~(\ref{upper_bound}) tends to zero in the continuum 
limit $W=\alpha L\to\infty$. From the definition of the Riemann integral, 
the second term converges to the integral 
\begin{equation} \label{second_term}
\lim_{W=\alpha L\to\infty}\frac{1}{L}\sum_{l\in\mathcal{L}}|u(x_{l}) - \tilde{u}(x_{l})|  
= \frac{1}{2}\int_{-1}^{1}|u(x)-\tilde{u}(x)|dx. 
\end{equation}
Thus, it is sufficient to prove that the RHS of (\ref{second_term}) tends 
to zero as $\alpha\to0$. 

For any $\epsilon>0$ and some $t_{0}\in\mathbb{R}$ in Lemma~\ref{stability}, 
we use the triangle inequality to obtain 
\begin{equation} \label{second_term_bound} 
\frac{1}{2}\int_{-1}^{1}|u(x)-\tilde{u}(x)|dx 
< \frac{1}{2}\int_{-1}^{1}|u(x)-\tilde{u}(x,t_{0})|dx + \epsilon. 
\end{equation} 
From Lemma~\ref{lemma4} and the definition of the initial condition for the 
differential system~(\ref{differential_system_u}), the first term on the 
upper bound~(\ref{second_term_bound}) converges to zero as $\alpha\to0$. 
\end{IEEEproof} 

%% Use \section* for acknowledgement
\section*{Acknowledgment}
The work of K.~Takeuchi was in part supported by the Grant-in-Aid for 
Young Scientists~(A) (No.~26709029) from JSPS, Japan.

%% References:
%% We recommend the usage of BibTeX:
%%
\bibliographystyle{IEEEtran}
\bibliography{IEEEabrv,kt-isit2014}
%\bibliographystyle{IEEEtran}
%\bibliography{definitions,bibliofile}
%%
%% where we here have assume the existence of the files
%% definitions.bib and bibliofile.bib.
%% BibTeX documentation can be obtained at:
%% http://www.ctan.org/tex-archive/biblio/bibtex/contrib/doc/
%%
%%
%%
%% Or manual references (pay attention to consistency!):
%%\begin{thebibliography}{1}
%%\bibitem{shannon1948}
%%  C.~E. Shannon, ``A mathematical theory of communication,''
%%  \emph{Bell System Techn. J.}, vol.~27, pp. 379--423 and 623--656,
%%  Jul. and Oct. 1948. 
%%\end{thebibliography}

\end{document}